\RequirePackage{fix-cm}
\documentclass{svjour3}                     
\smartqed  
\usepackage{graphicx}
\usepackage{epstopdf}
\usepackage{subfigure}
\usepackage{hyperref}
\usepackage{amsmath}
\begin{document}

\title{Flow visualization and PIV of wind-waves (2019 Edition)
}


\author{Chen Junwei
}


\institute{J. Chen \at
              State Key Laboratory for Turbulence and Complex System, College of Engineering, Peking University, Beijing, 100871, China \\
              \email{1601214787@pku.edu.cn}
}

\date{Received: date / Accepted: date}

\maketitle

\begin{abstract}
This article presents a flow visualization method for wind-waves, as well as a technique to measure flow field on two sides of interface by stereoscopic particle image velocimetry (PIV) simultaneously. The new flow visualization method applied a special illumination setup to enhance reflection and weaken refraction on interface, which improves contrast and detail performance of photos. The concatenation of flow visualization photo along flow direction is able to demonstrate the scenario from capillary waves to gravity-capillary waves at the early stage of wind-waves. After that, time-resolved stereoscopic PIV is employed in vertical planes on both sides of interface with two camera pairs. Two kinds of tracking particles with different scattering performance are spread in air or water separately. Then interface detection algorithm is designed to separate different particles and determine interface position. Adaptive min-max normalization and near-wall image preprocessing is used to improve near-wall and main stream PIV result. Finally, clip arts of results show the flow field in the same time in air and water near interface.

\end{abstract}

\section{Introduction}
\label{intro}
Wind-waves are common phenomenon in nature and attract a perpetual enthusiasm from fluid scientists. Over the past several decades, there are abundant literature on such phenomenon along with promotion by the evolved progress of techniques. The interface itself, the development of waves, flow on both sides of interface as well as their interaction is important in understanding wind-waves.

Flow visualization is the most direct method in wind-waves experiment, but its result contains even more details comparing to quantitative measurements, as there is no conversion from image to elevation field of interface. Some case of flow visualization is deduced in laboratory or natural environment, such as Banner and Phillips (1974), Su (1982), Kawamura and Toba (1988), Banner and Peirson (1998). J{\"a}hne et al. (1994) have introduced a stereo-photography method and then calculated slope of interface in field of view. Different approaches are applied to measure two-dimensional elevation field of air-water interface, such as intrusive or non-intrusive gauge array (Hammack et al. 1989; Bock and Hara 1995), background oriented schlieren (Moisy et al. 2009; Morris 2004; Fouras et al. 2008; Gomit et al. 2013), structured light (Tsubaki and Fujita 2005; Cobelli et al. 2009) and volume reconstruction of particles floating on the surface of water (Douxchamps 2005; Turney et al. 2009; Turney and Banerjee 2013; Aubourg et al. 2019). Some of these measurement are combined with PIV on the interface or on a horizontal plane near surface.

PIV is a widely used measurement method using plenty of tiny particles in fluid as tracers and calculate the velocity of particles within areas on the image. Many studies have improved the experimental arrangement and algorithm of PIV towards high-dimensional and more reliable results. Reul et al. (1999) measured air flow over a breaking wave. Siddiui et al. (2001) obtained current flow field with an additional profile camera to record interface, infrared visualization is executed in the same time. Misra et al. (2006) and Dussol et al. (2016) detected interface position in PIV image via the distribution of particles, while Andre and Bardet(2014), Buckley and Veron (2017), Vollestad et al. (2019) via laser-induced fluorescence. Young and Hyung (2011), Park et al. (2015) found interface by matching texture to photo, which is only feasible in the situation of quasi-one-dimensional interface. Actually, Buckley and Veron (2017) has listed nearly 60 articles about PIV of wind-waves. However, flow measurements near an arbitrary interface are still challenging for several reasons. The location of interface position is a pivital problem as regard to the correctness of results. Besides, high shear region near interface requires a robust algorithm. Furthermore, the volatile interface generates strong reflection light and nonuniform refraction light that will deteriorate the image quality in experiments.

Since surface gravity waves strongly couple to adjacent winds and currents (Sullivan and McWilliams 2010), it is a natural idea to measure flow field on both sides of interface simultaneously. This paper presents a sketch of applying stereoscopic PIV in wind and current near the interface of wind-waves with two camera pairs working concurrently, which includes experimental setup, interface detection and appropriate PIV algorithm. Besides, a simple but effective method is introduced, which produces high-contrast and more detailed flow visualization of interface. The paper is organized in the following way. Flow field setup and visualization method is given in Sect. 2. PIV image acquisition (Sect. 3) and processing (Sect. 4) follow, the latter one includes interface detection procedure and PIV algorithm special for experimental case in this article. Finally the result and discussion are in Sect. 5.
\section{Flow visualization}
\label{sec:1}
\subsection{Flow field setup}
\label{sec:2}
Wind waves are generated in the wind-water tunnel at Peking University. This wind-wave tunnel is a combination of the low-turbulent water channel and an elementary wind tunnel. The close-loop water channel has two test sections and the second one measures 6 m x 40 cm x 40 cm. The wind tunnel is mounted over water tunnel and wind begins to interact with water from one metre at the second test section. A 40 cm x 30 cm honeycomb filter at the outlet of wind tunnel straightens the wind. Wind speed varies from 0 to 6.0 m/s and water speed is up to 0.60 m/s. As wind accelerates water near surface but the total flow of water is controlled by the console of water tunnel, there will be backward flow at bottom region when total flow is insufficient. In most cases of our experiments, water speed is adapted to wind speed to make the bottom region of water nearly stall.
\subsection{Visualization method}
\label{sec:3}
\begin{figure}
\includegraphics[width=0.5\textwidth]{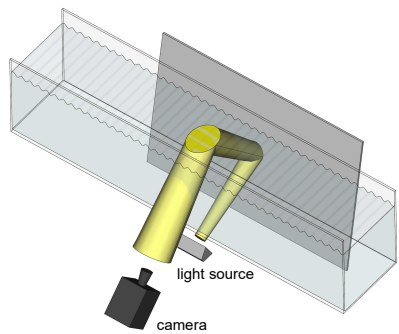}
\caption{Schematic of flow visualization}
\label{fig:1}
\end{figure}
The plane-view visualization of interface was usually snatched from the air side in previous articles. And the line-of-sight is tilt from vertical to make photo become more stereoscopic with fair contrast.

However, when watching up from the underwater side, the contrast of image will be significantly improved. Comparing to refraction, reflection is more sensitive to small undulation of liquid surface because of its higher deflection angle. On the other side, refractive index of water is larger than that of air. When the camera is filming interface from somewhere above the interface, part of light below the interface will refracts there and be recorded whatever the tilt angle of line-of-sight is. In contrast, if camera were below the interface and watched upwards, the energy ratio of reflection to refraction will increase dramatically as the incident angle approaches to that of total reflection, i.e. the camera will record mostly reflection light and the contrast of photo is improved. This can be explained by reflectivity- or transmissivity-power-ratio derived from Fresnel equations.

Fig. \ref{fig:1} shows a schematic diagram of the experimental arrangement for flow visualization. A screen covered by white paper sits in one side of wind-water tunnel and faces toward the object interface. A sodium-vapour lamp illuminates the screen where light diffused reflects. A PCO sCMOS camera (16 bit grayscale, 2560 x 2160 $pixel^2$ at 50 fps) is in the other side of wind-water tunnel. The field of view (FOV) is set to be approximately 100 cm x 40 cm using a 28 mm lens and 20 cm x 40 cm using a 60 mm lens. The camera has a pitch angle about $45^{\circ}$ upwards to improve the contrast of image as mentioned above. LaVision v4.2 Scheimpflug adapter is used and the aperture of lens is properly reduced to help the image system focus on the interface.

A bi-lineal mapping function is used to correct the perspective distortion. Several points are marked with their position both in raw image and real world to fit the mapping function, and then a trapezoidal shape region of raw image is dewarped into a rectangle shape image with uniform scale.

\begin{figure*}
\subfigure[]{
\includegraphics[width=.24\textwidth]{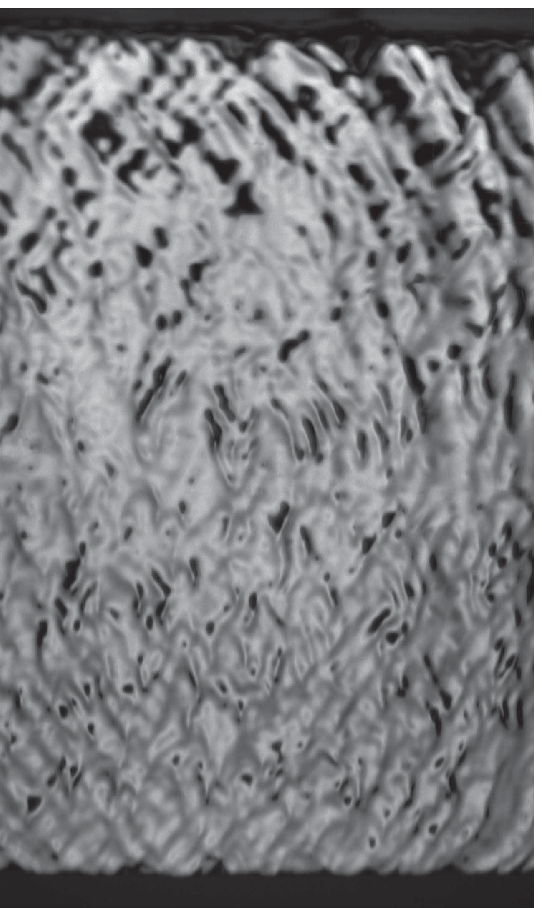}}
\subfigure[]{
\includegraphics[width=.24\textwidth]{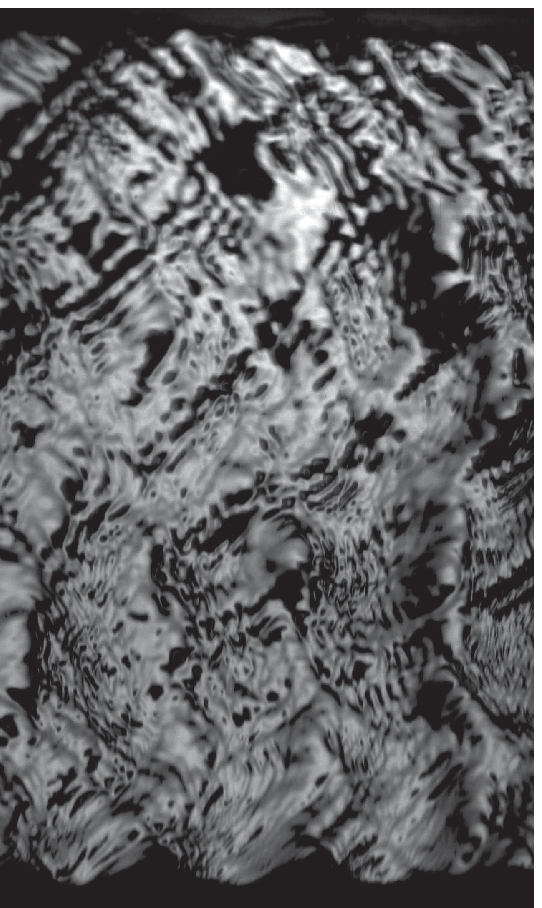}}
\subfigure[]{
\includegraphics[width=.24\textwidth]{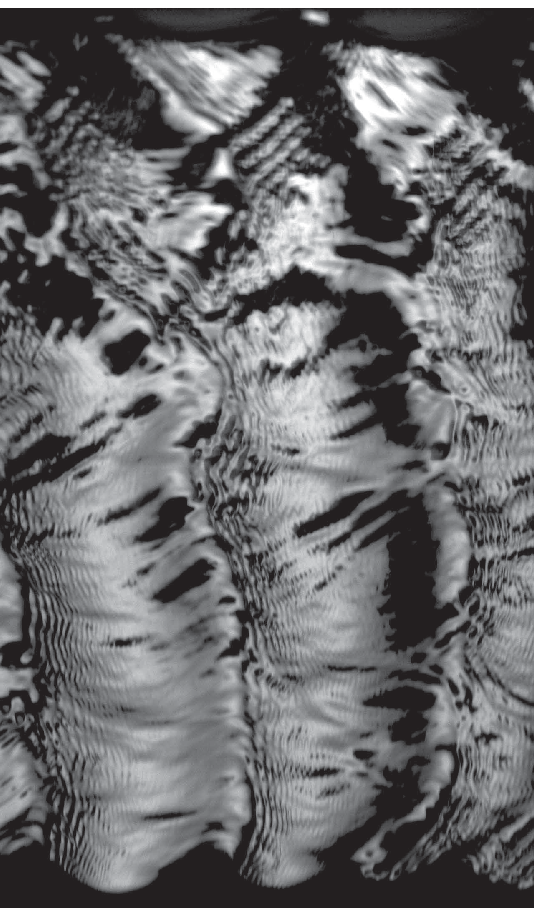}}
\subfigure[]{
\includegraphics[width=.24\textwidth]{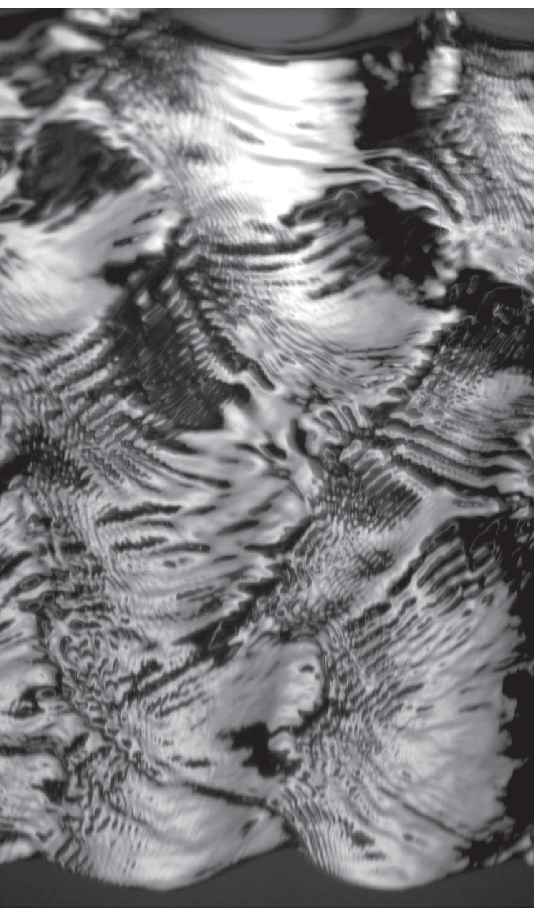}}
\caption{Flow visualization at 260cm from the downstream of wind tunnel. The flow is from left to right and the width of flow field is 40cm. The Wind speed is \textbf{a} 3.1m/s; \textbf{b} 5.0m/s; \textbf{c} 5.2m/s; \textbf{d} 6.0m/s.}
\label{fig:2}
\end{figure*}

\begin{figure*}
\includegraphics[width=\textwidth]{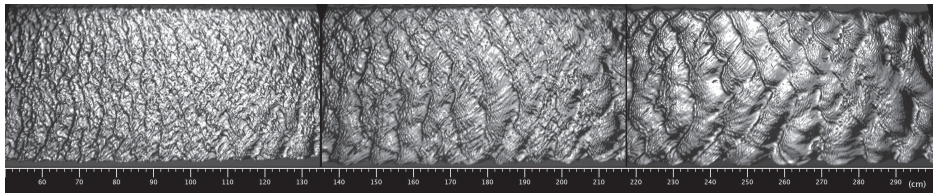}
\caption{Flow visualization of interface from 50cm to 300cm on the downstream of the exit of wind tunnel. Independent measurements over three successive sections are concatenated together. The flow is from left to right.}
\label{fig:3}
\end{figure*}

Fig. \ref{fig:2} shows interface under different wind speed. The FOV is about 20 cm x 40 cm. Flow is from left to right, and the upper and lower margin of flow field happen to show the height variation of interface. Wind waves are changing from capillary waves to gravity-capillary waves when wind speed increases. Many tiny wavelets in the downstream of the wave crests can be seen clearly, which is called 'incipient breaking' by Phillips and Banner (1974). In order to show the transition process from capillary waves to gravity-capillary waves at the early stage of wind-waves, three successive sections of interface visualization measured independently are concatenated together. Fig. \ref{fig:3} demonstrate interface between 50 cm and 300 cm from beginning. The flow is from left to right. Also, perspective distortion is corrected.
\section{PIV image acquisition}
\label{sec:4}
\begin{figure*}
\includegraphics[width=\textwidth]{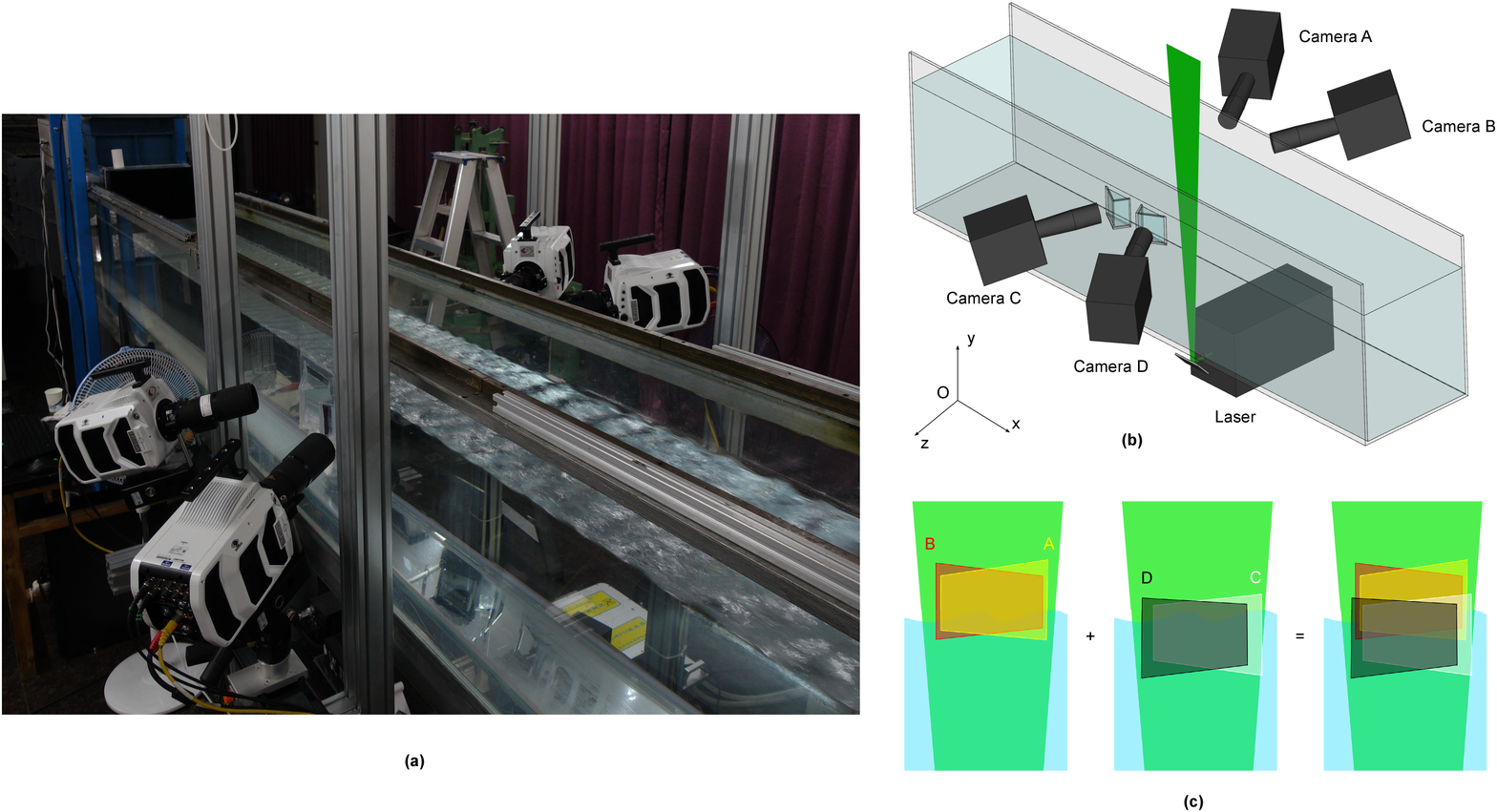}
\caption{Camera and laser setup for PIV and FOV of each camera. Air and water is flowing from left to right in all pictures. \textbf{a} Presentation of experimental setup. \textbf{b} A 3D sketch of imaging system. \textbf{c} field of view of each camera. Perspective distortion is a little exaggerated. Notice that image taken by camera A and B will turn over vertically to make the flow from left to right.}
\label{fig:4}
\end{figure*}

Stereoscopic PIV is employed simultaneously in air and water. Four Phantom v2512 (12 bit grayscale, 25600 fps at full resolution 1280 x 800 $pixel^2$, 72 GB internal RAM) fast-speed cameras are used to grab particle images in air and water. A ZKLASER DPQ-527-60PIV-DF Nd:YLF laser shades 1 mm light sheet vertically upwards from the bottom of wind-water tunnel. The energy of laser is 6 mJ per pulse at 10000 Hz. As Fig. \ref{fig:4} shows, camera A and B focus on flow field above the interface while C and D focus on that below the interface. The cameras have about 10$^{\circ}$ pitch angle to surmount the relief of interface beyond the focusing plane so that near surface region can be snatched without obstacle. The angle between two cameras in each group is about 60$^{\circ}$ to balance stereo-PIV error in three direction (Prasad 2000; Bhattacharya et al. 2016). 200 mm lens are mounted on Scheimpflug adapter and fields of view are set to be approximately 5 cm x 3 cm on each camera and the interface always traverses the picture. Laser and all cameras follow the same time sequence generated by LaVision PTU. Sampling rate is from 8000 fps to 15000 fps, which is adapted to wind speed. Besides, two acrylic prisms filled with water are stuck on the lateral face of wind-water tunnel. The prisms have a 30$^{\circ}$ edge. In this way, the scattered light of particles in water will have small incident angle as far as possible when it refracts to air and reach camera aftermath, so that the blur of each particle spot will be reduced (Prasad and Jensen 1995). The aperture for camera A and B is 1/4 and for C and D is 1/5.6. Each case takes 10000 frames during experiment.

Two kinds of particles are added into air and water. The choice of particles is under synthetical consideration among following performance, light intensity of spots in image, difference of two kinds of particles to be separated and less impact on wind-waves while floating on interface. Hollow acrylic particles, 11 $\mu$m in mean diameter, from Potter Industries are introduced into water. Again, this particles are poured into water very gently along streamwise to supply more particles just beneath the interface. Fume composed of particles, about 1-3 $\mu$m in diameter, is injected at the upstream of wind tunnel by a server fan.

The calibration of two camera groups is done separately on LaVision DaVis 10. A LaVision 058-5 calibration plate, which has 2 level on each side and 5 mm between two neighbour spots, mounts on a dock, where it can only move upward or downward if the dock were fixed. The first level of calibration plate coincides with light sheet and the height of calibration plate is set to a proper position. Each camera takes 100 pictures of calibration plate and calculates average intensity of pictures to reduce noise. The calibration generates 3rd-order polynomial functions for mapping things between raw image and real world. The scale factor above the interface is about 28.0 pixel/mm and below the interface is about 28.5 pixel/mm in images corrected into real world. To improve calibration quality, water level is changed temporally so that more points are recorded. Also, photos for self-calibration are grabbed. Particles in air are replaced with about 10-$\mu$m water droplets from Eau Thermale, in lower particle density.

\begin{figure*}
\subfigure[]{
\includegraphics[width=.5\textwidth]{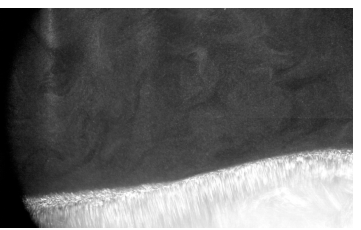}}
\subfigure[]{
\includegraphics[width=.5\textwidth]{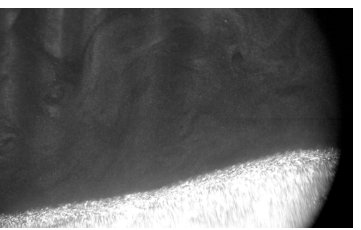}}
\subfigure[]{
\includegraphics[width=.5\textwidth]{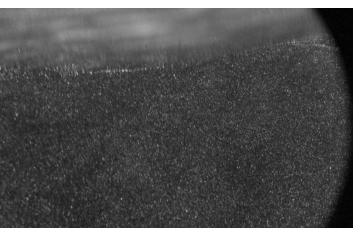}}
\subfigure[]{
\includegraphics[width=.5\textwidth]{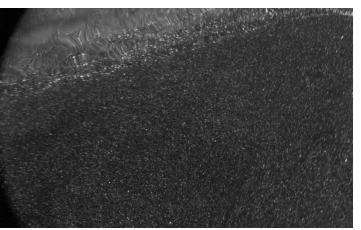}}
\caption{Images taken by four cameras at the same time. \textbf{a} to \textbf{d} are images taken by camera A to D.}
\label{fig:5}
\end{figure*}

Photos taken by four cameras in the same time are shown in Fig. \ref{fig:5}. Fig. \ref{fig:5}a and b are particle images of air flow while c and b are of water flow. Fig. \ref{fig:5}a and b have been turned over vertically because the cameras are in the other side of wind-water tunnel. Concave black shading occurs in the left side or right side of pictures because the lens have tilting angle up to 10$^{\circ}$ to meet Scheimpflug criterion and the 42 mm CMOS chip is too large for the 200 mm lens used. Nonetheless, such shading has little influence on stereo PIV calculation for the FOV of two cameras in one group may not overlap in some edge area hence velocity will not be calculated there.

\section{PIV Image processing}
\label{sec:5}
The algorithm of interface detection and PIV is illustrated schematically in Fig. \ref{fig:6}. Interface detection includes four steps while PIV includes three steps. There is a sequential order between interface detection and PIV, because PIV image preprocessing will use interface data.

\begin{figure*}
\subfigure[]{
\includegraphics[width=.5\textwidth]{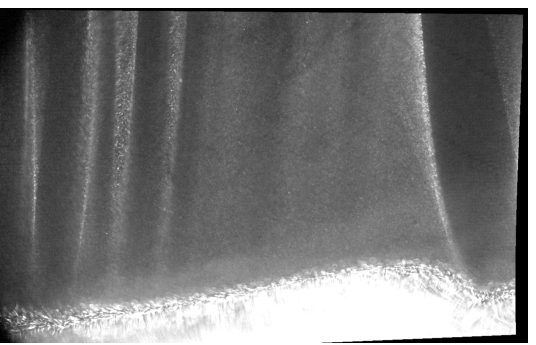}}
\subfigure[]{
\includegraphics[width=.5\textwidth]{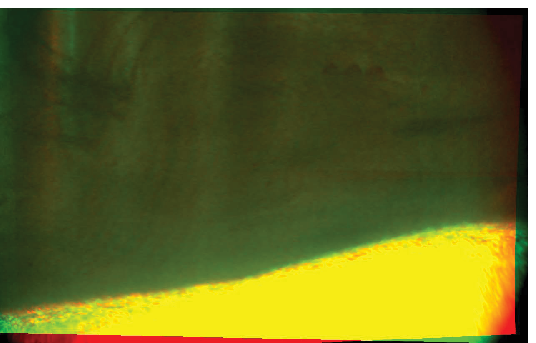}}
\subfigure[]{
\includegraphics[width=.5\textwidth]{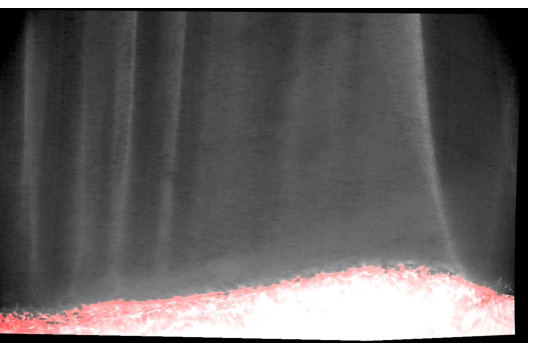}}
\subfigure[]{
\includegraphics[width=.5\textwidth]{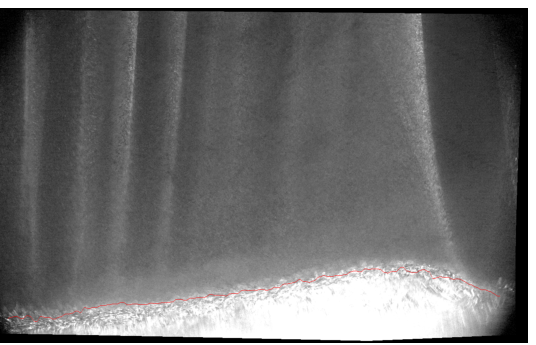}}
\caption{Procedure of interface detection before spatial and temporal smoothing, the flow is from left to right. \textbf{a} Original pictures to be processed are showing in pseudo-colour form, Images grabbed by camera A and B is turned over vertically and perspective distortion is corrected. \textbf{b} Pictures after preprocessing for interface detection. \textbf{c} After comparing with threshold, 'true' pixels are highlighted. \textbf{d} Interface position is detected and preliminary smoothed, then showed in red or green line.}
\label{fig:7}
\end{figure*}

It would be better to do self-calibration before processing. For calibration in stereoscopic PIV, apart from velocity calculation errors (Wieneke 2005), near-wall results may be incorrect for the non-coincidence of interface in wind-waves experiments. The easiest way to inspect calibration quality is to show stereoscopic image pair in a single picture, for instance, putting image taken by camera A in red channel and image taken by camera B in green channel of a pseudo colour picture, see Fig. \ref{fig:7}a, then observe whether interface in red and green channel coincide or not.
\subsection{Interface detection}
\label{sec:6}
Before interface detection, the content in picture of each camera should be clarified first. There are two kinds of particles in the picture focusing on air flow, fume in air and acrylic particle in water, acrylic particles are much brighter than fume because their diameter is larger. There is only acrylic particles on the picture focusing on water flow, scatter light of particles directly recorded by camera spread below the interface while that reflected by the interface almost appear above the interface. The scattering of particles follows Mie scattering law (Raffel et al. 2018). As light shades upward in the experiment setup mentioned above, light scattering oblique upward is generally more powerful than that scattering oblique downward. This explains why particles image focusing on water flow has different light intensity on two sides of interface.

Edge detection is frequently utilized in wall and free surface detection, in which drastic change of light intensity, such as profile of particles and solid wall, is found in images. However, particles spread on both side of interface in our case. It is difficult to recognize interface from edge detection result. Instead of that, threshold is applied by comparing to images as two kinds of particles have different light intensity and size in picture of air flow. The process of first frame is a little different to successive frames with respect to time-series enhancement, but they are similar on the whole. There are generally four steps in interface detection, and the first to the third step are operated on corrected images about air flow.
\subsubsection{Image preprocessing}
\label{sec:7}
As acrylic particles in water are much bigger than fume particles in air and their scattering direction is more forward, acrylic particle spots are about ten times brighter than fume particles as well as occupy morn pixels in the image taken by camera A and B. Besides, for wind waves in early stage, water is not much accelerated, wind flow is much faster than water flow. If the sample rate were adapted to airflow, particles in water are nearly stall between two adjacent frames. It is recommended to calculate time-series sliding minimum for every three images, in this way, particles in air are weakened while particles in water are almost preserved.

Then, due to different sizes of two kinds of particles in image, the grayscale morphological opening, which consists of grayscale morphological erosion and dilation in sequence, is applied to particle image. Erosion is to eliminate the edge of bright area by seeking the dimmest pixel in the neighbourhood of one pixel and then replacing the latter. Besides, dilation does the contrast way and extend bright area. If the parameter of grayscale morphological opening is proper, fume particles will be removed in erosion as far as possible, while acrylic particles will be reduced in erosion and then recovered in dilation. Image before and after preprocessing is shown in Fig. \ref{fig:7}a and \ref{fig:7}b.
\subsubsection{Interface detection via local threshold}
\label{sec:8}
After pre-processing, images are compared with threshold to separate region contain acrylic particles from region contain fume particles, then it yields binary images. As mentioned above, some margin area of images may be dim. A local threshold set is applied to this situation, i.e. the threshold near dim area is lower than that in ordinary area. However, if a smaller camera sensor or a lens with larger aperture were used, a constant threshold is adequate.

Morphological opening is applied to binary image, which is to clear sporadic 'true' pixels above the interface, the result is shown as highlight area in Fig. \ref{fig:7}c. Aftermath, row position is determined by seeking the first 'true' pixel in every column. The raw position is not smooth due to discreteness of particle image, so the preliminary smoothing, which contains median and Gaussian filter, is applied before temporal and spatial smoothing in the next step, the result is shown in Fig. \ref{fig:7}d.

There is an additional procedure in binary image processing special for the successors of first frame. Upper margin of 'true' pixels is limited by the interface in previous frame, hence can not exceed it too much. To make the algorithm more robust, it is recommended to set limitation alternately, i.e. interface on camera A is the source of limitation in next frame on camera B, and vise versa. The reason is bright taint above interface caused by conglomeration of fume particles, reflected light from interface, off-plane particles, converging and diverging light passed through interface are less probable to arise in the same place on two cameras simultaneously.
\begin{figure}
\includegraphics[width=0.5\textwidth]{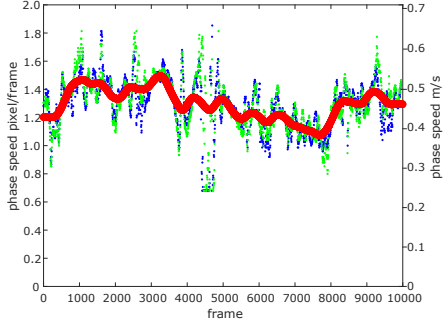}
\caption{Phase speed of interface. Blue dots represent phase speed detected from camera A and Green dots from camera B. Red line is phase speed after smoothing.}
\label{fig:8}
\end{figure}
\subsubsection{Spatial and temporal smoothing}
\label{sec:9}
In this step, interface position is smoothed using data in several vicinity frames. Albeit the interface is changing all the time, its shape is nearly frozen when moves downstream so that phase-locked averaging can be applied to smooth the interface. Phase speed of waves should be calculated before averaging. It is calculated by finding correlation peak of interface position between two frames. As the phase speed is about 1.2 pixel/frame, see Fig. \ref{fig:8}, numerical error will be significant if two adjacent frames were used. The gap is chosen from 25 frames to 250 frames. Due to many spikes in the series of displacement, median filter and Gaussian filter are applied to smooth the data.

\begin{figure}
\subfigure[]{
\includegraphics[width=0.5\textwidth]{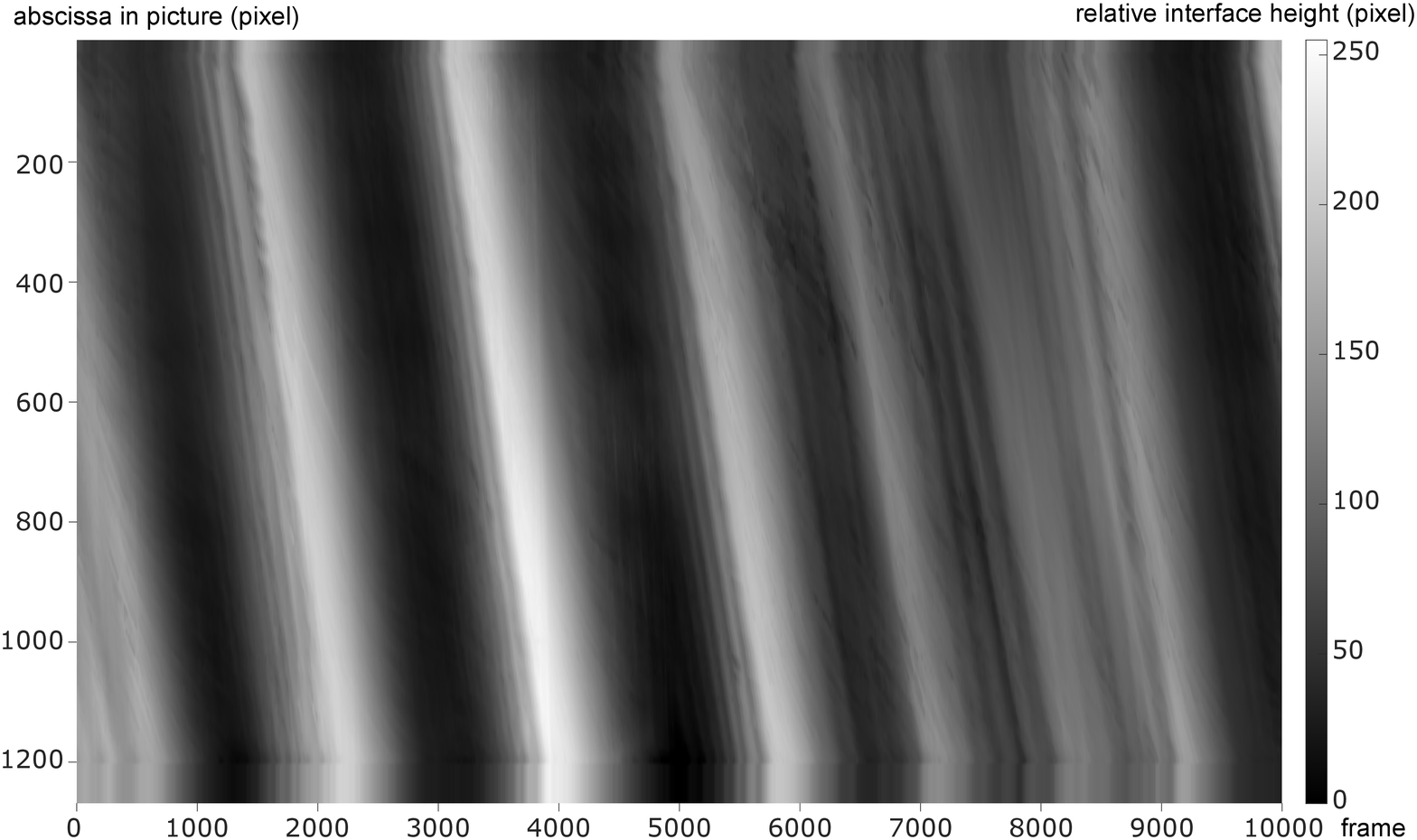}}
\subfigure[]{
\includegraphics[width=0.5\textwidth]{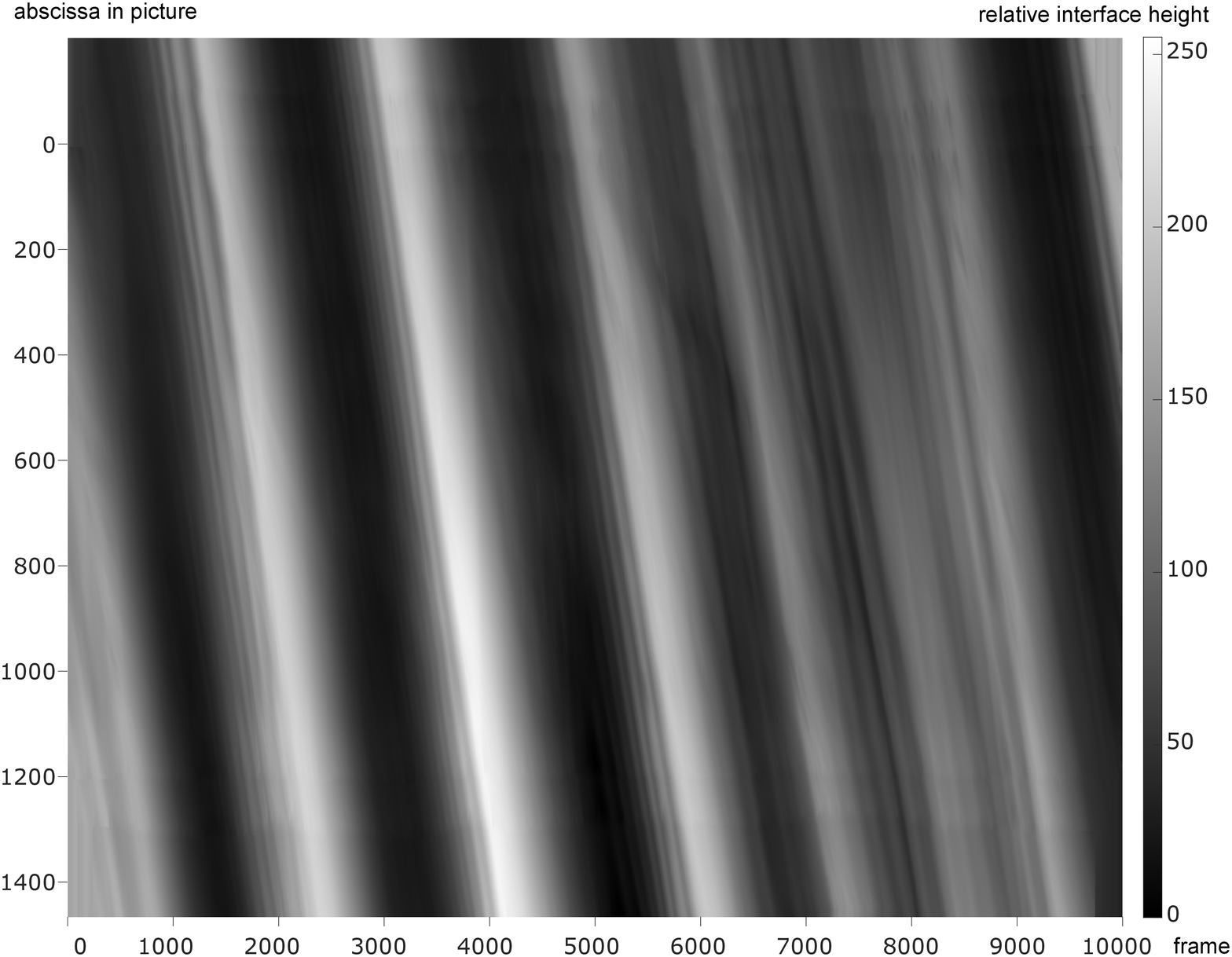}}
\caption{Relative interface height \textbf{a} before and \textbf{b} after spatial and temporal smoothing, greyscale represent relative interface height. The interface is extended by 200 pixels in two sides as well as smoothing.}
\label{fig:9}
\end{figure}

For the interface position series $\boldsymbol{s}(\boldsymbol{x}, t)$, where $\boldsymbol{x}$ are series of x-coordinate, t is time and every element in $\boldsymbol{s}$ corresponding to element in $\boldsymbol{x}$ at the same place. Let phase speed to be $\boldsymbol{v}_\phi(t)$ and the time interval to be $\Delta$t, than interface moves $\Delta\boldsymbol{x}(t) \approx \boldsymbol{v}_\phi(t){\Delta}t$ in a time interval. If 2N + 1 frames were used to smooth the interface, and suppose that phase speed didn't change in short time, then smoothed interface in $t_0$ is
\begin{equation}
\boldsymbol{s}_e(\boldsymbol{x}_e, t_0) = \overline{\boldsymbol{s}(\boldsymbol{x} + n\Delta\boldsymbol{x}(t_0), t_0-n\Delta t)},
\end{equation}
where $\boldsymbol{x}_e$ is extension of $\boldsymbol{x}$ to the sides, $\boldsymbol{s}_e$ is interface position on $\boldsymbol{x}_e$, and n is from -N to N. The averaging will omit terms without value. Since interface position is discredited and averaging is conducted at every elements in $\boldsymbol{x}_e$, one-dimensional interpolation has to be done before. Interface position in all frames before and after phase-locked averaging is shown in Fig. \ref{fig:9}, where higher brightness means relatively higher position. The interface is more smooth and extended to two sides by 200 pixels in Fig. \ref{fig:9}b. There are many tiny wavelets behind wave crest, which can be seen in Fig. \ref{fig:2} too, but some small amplitude wavelets are still smeared out.
\subsubsection{Applying interface position to all cameras}
\label{sec:10}
In step 1-3, the interface in images about air flow have been detected. In this step, the detected interface are applied to images about water flow. Also, self-calibration and perspective correction are required first. As the scale factor for two calibration system on other side of interface is not always the same, the interface series detected should be amplified or reduced to adjust the scale factor of calibration for water flow. Then move the interface to the borderline in picture, where particle intensity is different on two sides.

\begin{figure}
\subfigure[]{
\includegraphics[width=0.5\textwidth]{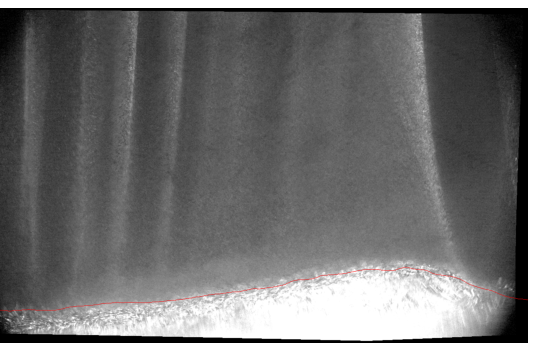}}
\subfigure[]{
\includegraphics[width=0.5\textwidth]{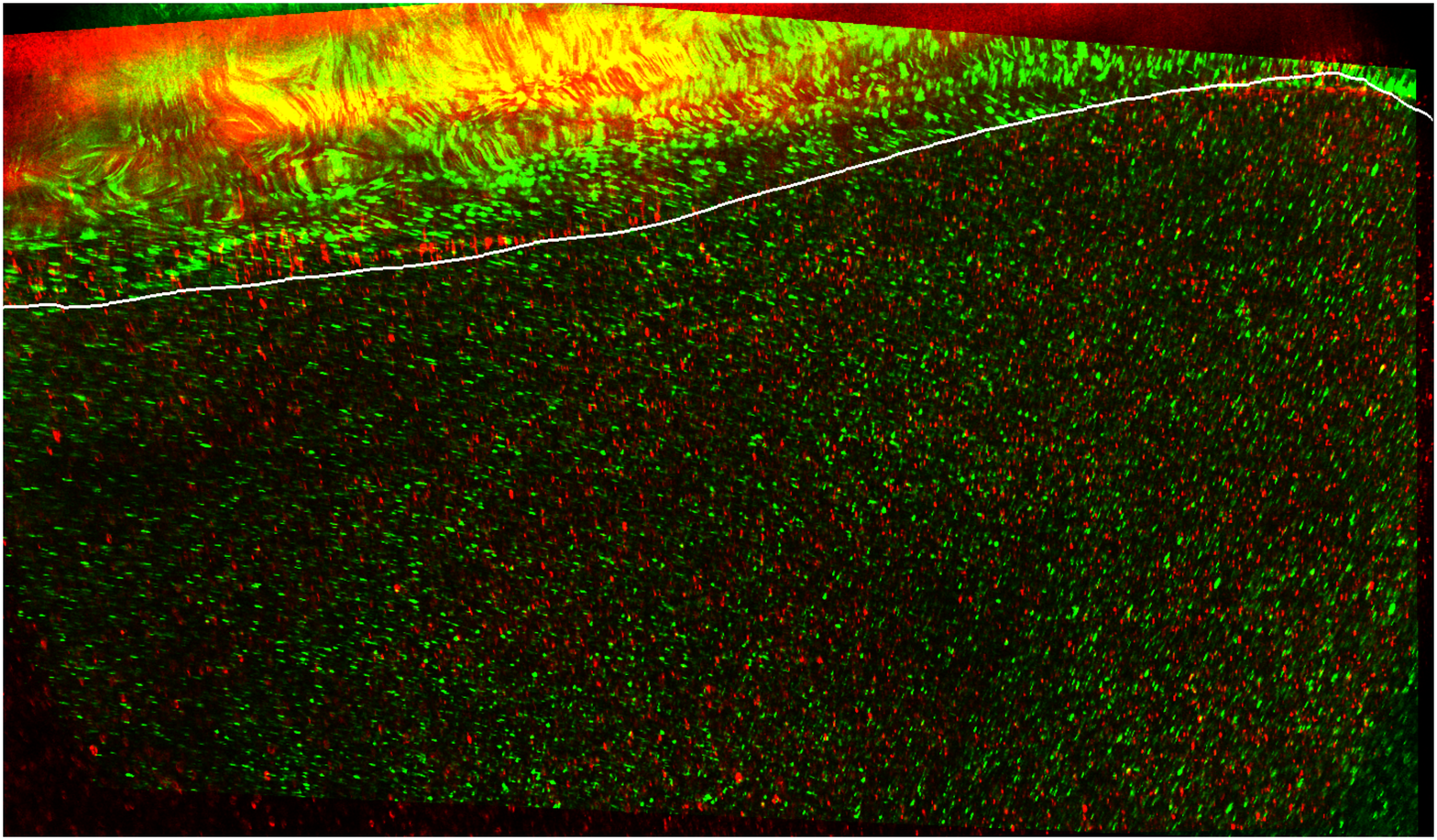}}
\subfigure[]{
\includegraphics[width=0.5\textwidth]{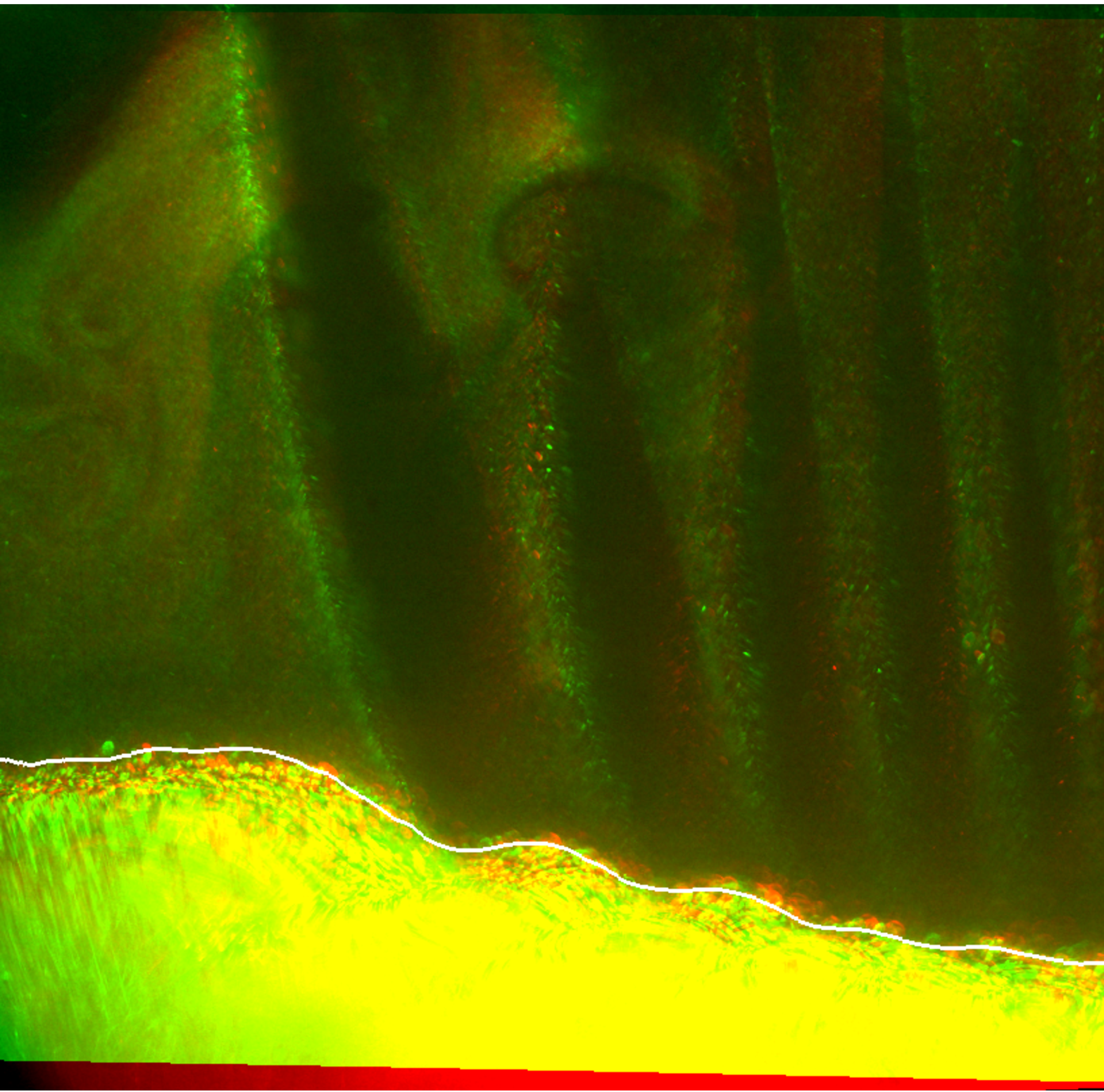}}
\subfigure[]{
\includegraphics[width=0.5\textwidth]{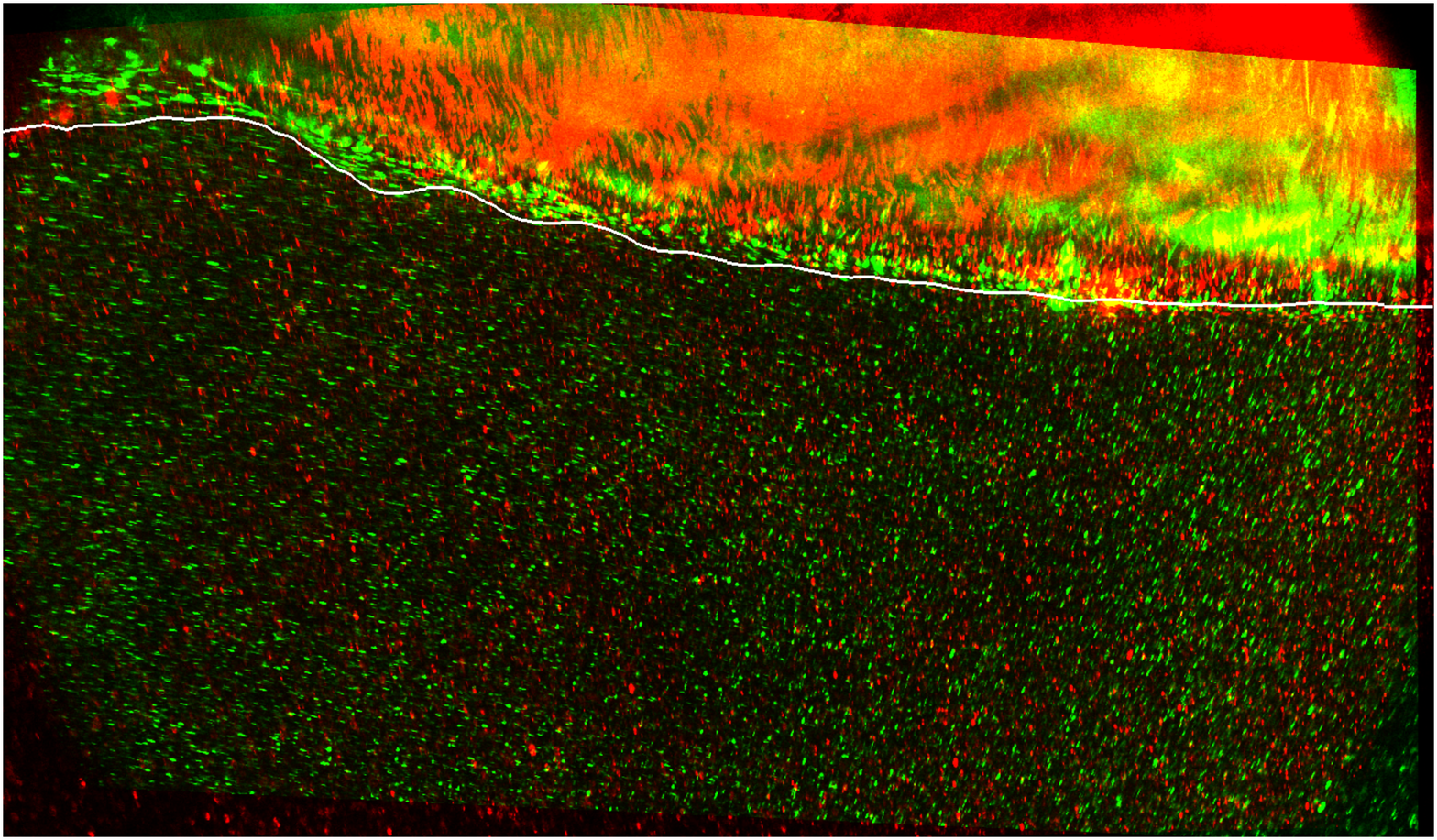}}
\caption{Interface detection results, image of \textbf{a} and \textbf{b} are taken in at the same time, while \textbf{c} and \textbf{d} are taken in another time. Flow is from left to right in all images and white line represent interface detected.}
\label{fig:10}
\end{figure}

Fig. \ref{fig:10} shows two scenery of interface detection result, and it works stably in both upstream and downstream of wave crests. The accuracy of interface detection method mentioned above is about 5 pixels in our experimental case. This method is a bit time-consuming for its multiple image morphological processing and interpolation, because the algorithm must be stable enough to deal with thousands of frames. The processing time for 10000 frames is about 3 hours on a quad-core platform via MATLAB.
\subsection{PIV processing}
\label{sec:11}
Adaptive min-max normalization is applied to images taken by camera A and B. Assume that the light intensity distribution of an image, I(x, y) can be characterized by two components: (1)source illumination incident on the scene being viewed, and (2)illumination reflected by the objects in the scene (Gonzales and Woods 2007). For particle images with noise, intensity distribution can be decomposed by
\begin{equation}
I(x, y) = I_{illu}(x, y)I_{part}(x, y) + I_{noise}(x, y),
\label{eq:2}
\end{equation}
where $I_{illu}$ is distribution of illumination, $I_{part}$ is reflectance by tracking particles and $I_{noise}$ is noise. The ideal situation is $I_{illu}$ is uniform and $I_{part}$ is homogeneous. Quality of image is deteriorated if either of them were dissatisfied. For non-uniform illumination, there are plenty of methods to rescue the image, such as min-max normalization introduced by Westerweel (1993). Adaptive min-max normalization combines two min-max normalization results with different filter size together. Suppose that normalization result of raw image by filter size n is $f_n\{I\}$, and $\widetilde{I}$ is multiply smoothed image, in which undulance of particles can not been seen. Then adaptive min-max normalization is
\begin{equation}
f_{a, b}^\star =
\left\{\begin{matrix}
 f_a\{I\}, \quad \sigma(\widetilde{I}) > H,\\
 f_b\{I\}, \quad \sigma(\widetilde{I}) < H,
\end{matrix}\right.
\end{equation}
where $a < b$, $\sigma(\widetilde{I})$ is standard deviation of $\widetilde{I}$ of every pixel with its several vinicity pixels, and H is threshold.

\begin{figure*}
\subfigure[]{
\includegraphics[width=0.5\textwidth]{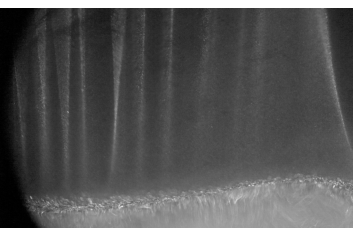}}
\subfigure[]{
\includegraphics[width=0.5\textwidth]{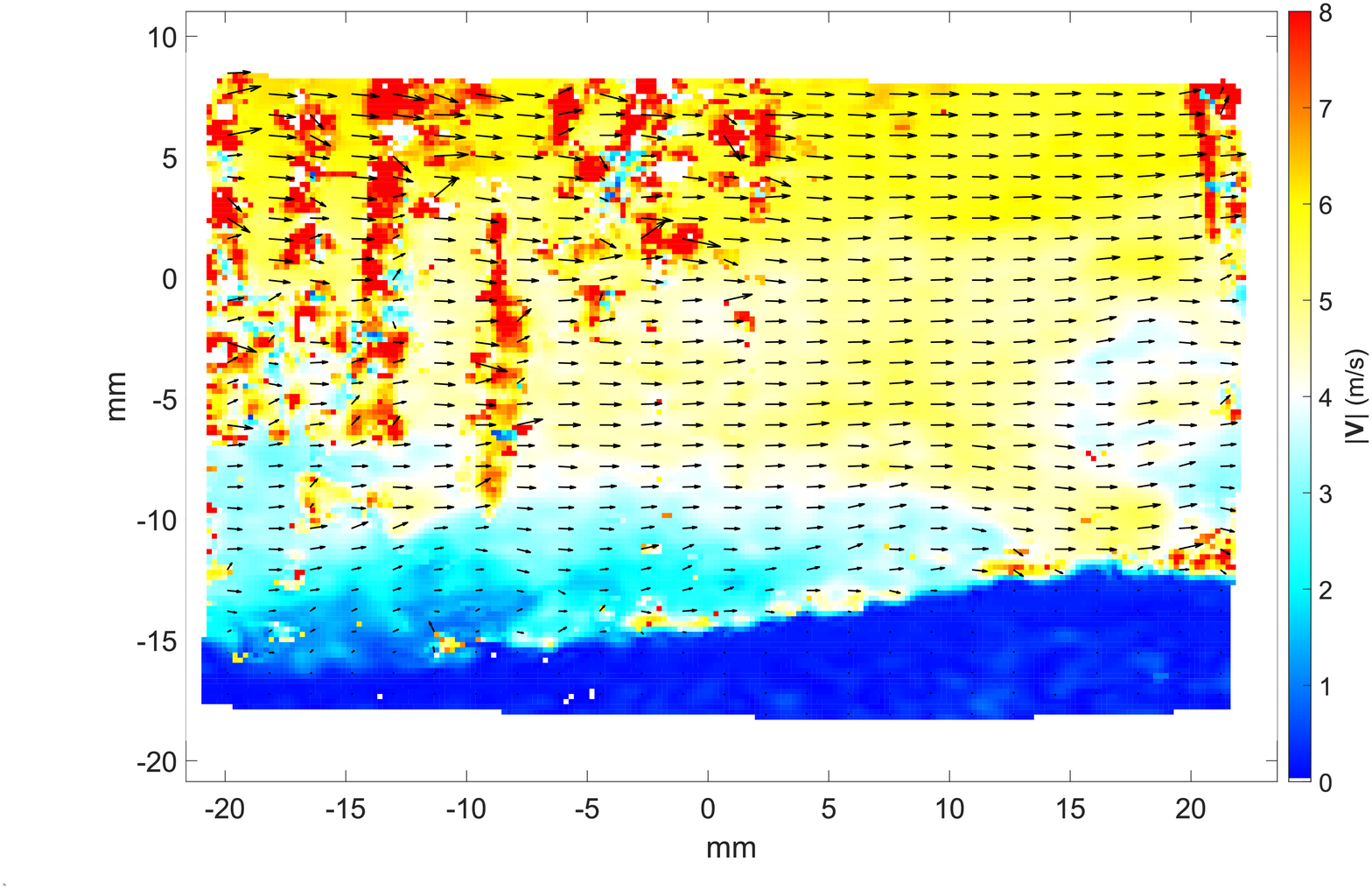}}
\subfigure[]{
\includegraphics[width=0.5\textwidth]{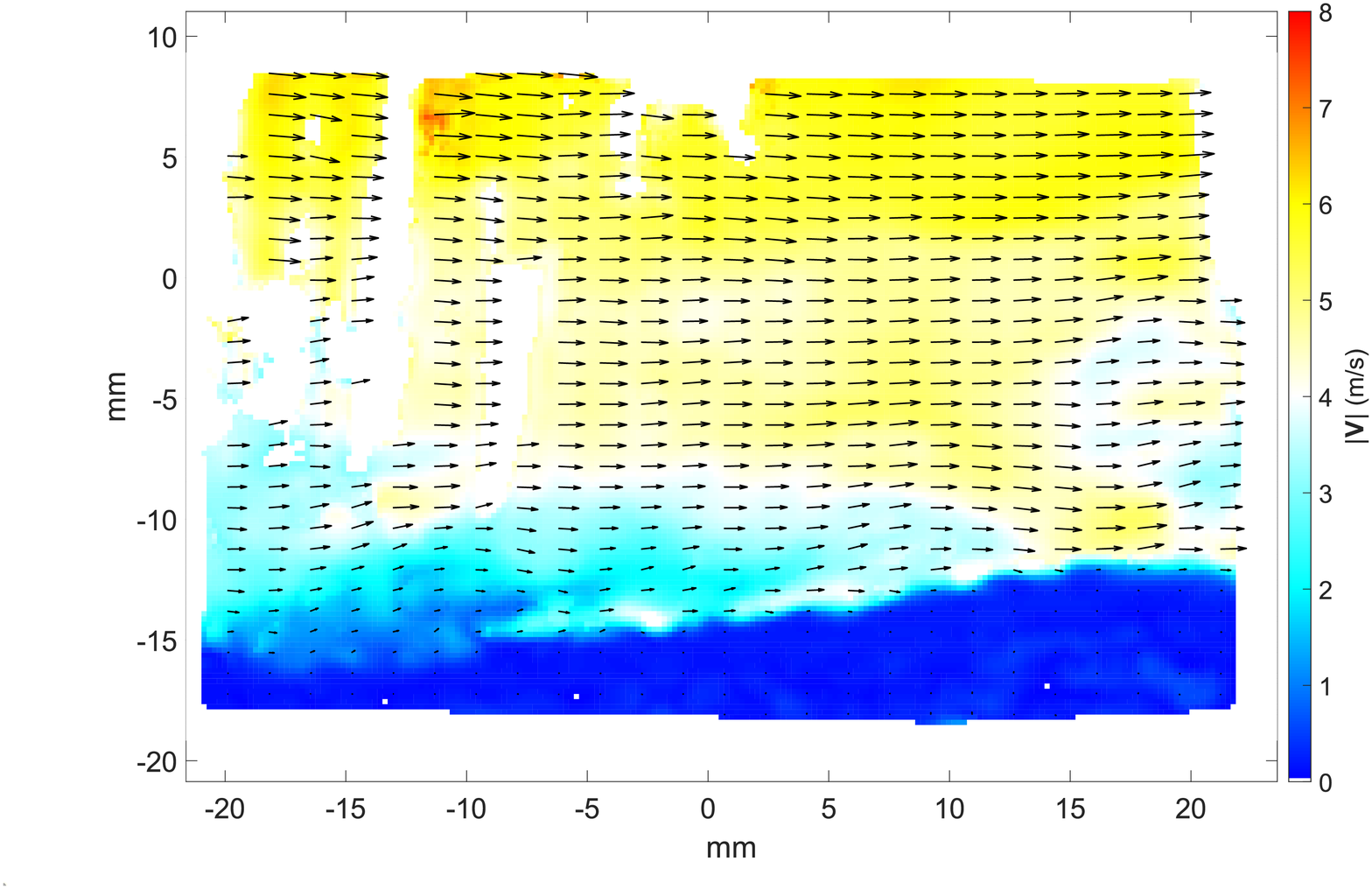}}
\caption{Correlation result are compared from images without or with adaptive min-max normalization. \textbf{a} is origin image of stereoscopic pair, while is \textbf{b} is the adaptive min-max normalization result of \textbf{a}. \textbf{c} is correlation result without image preprocessing, and \textbf{d} is correlation result of method in Sect. 4.2.}
\label{fig:11}
\end{figure*}
In classical PIV measurement, particle images are divided into a series of interrogation windows, and the particle distribution in interrogation window is one of the factors influencing vector calculation. Simply mask out particles on the other side of interface may not be a good precessing method, because it will cause a typical problem, lack or even absence of particles on other side in the interrogation window across the interface (Park et al. 2015; Theunissen et al. 2008). Optimal synthetic particles is applied to fill the other side of fixed or translating wall in order to improve velocity prediction in high-shear boundary layer (Zhu et al. 2013; Jia et al. 2017). Due to particles spreading in both phase and inclined camera mounting in the experimental setup in this article, particles are filled in both sides of interface in image, as mentioned above. The way for near-wall velocity calculation is to keep particles in other side of interface and treat them as optimal synthetic particles. Particles below the interface in images about air flow are simply kept after adaptive min-max normalization. Reflected particles above the interface in images about water flow act like mirror particles, they have same tangential velocity and opposite normal speed to their original part. Such mirror particles are processed as SP (Stationary particles) mode (Zhu et al. 2013), their intensity is reduced to one-third of the original.

The velocity vectors are obtained on LaVision DaVis 10 using multi-path cross-correlations and the final interrogation window is comparable to a 24 x 24 square window whose shape and weight is elliptical Gaussian with lateral-to-vertical factor 2:1. 75\% overlap is applied to improve vector field resolution, and vector spacing in air flow measurement is 214 $\mu$m while in water flow measurement is 211 $\mu$m. In vector post-processing, removal-and-replacement median filter and velocity range limitation are applied to remove outliers. The empty space after post-processing is filled with calculation result from time-resolved PIV sliding sum-of-correlation (Scarano and Moore 2012; Sciacchitano et al. 2012) in 5 frames with the same correlation setup and post-processing procedure as the first round. Due to water flow is much slower, velocity is calculated at intervals of 5 frames.
\begin{figure*}
\subfigure[]{
\includegraphics[width=0.5\textwidth]{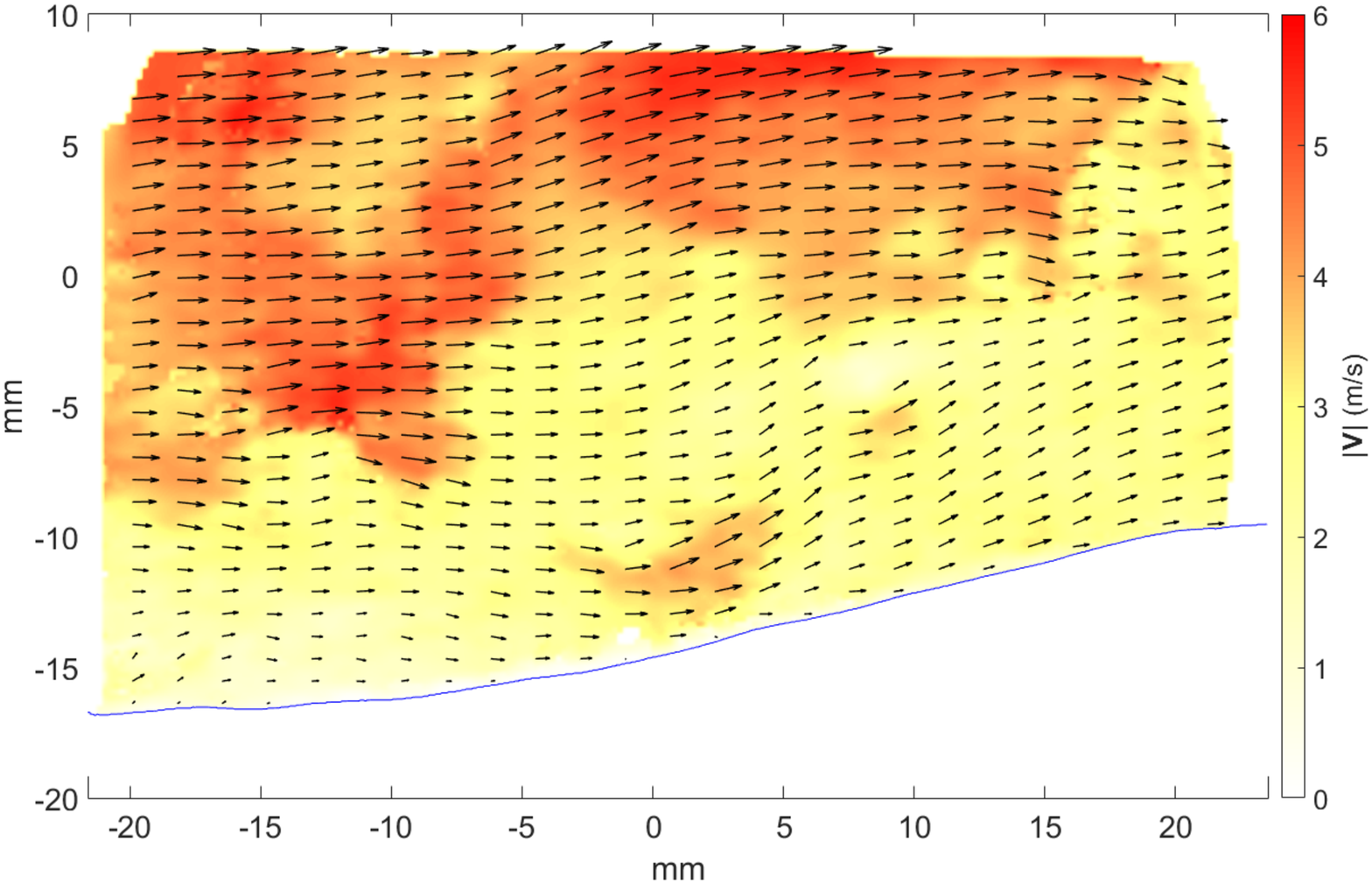}}
\subfigure[]{
\includegraphics[width=0.5\textwidth]{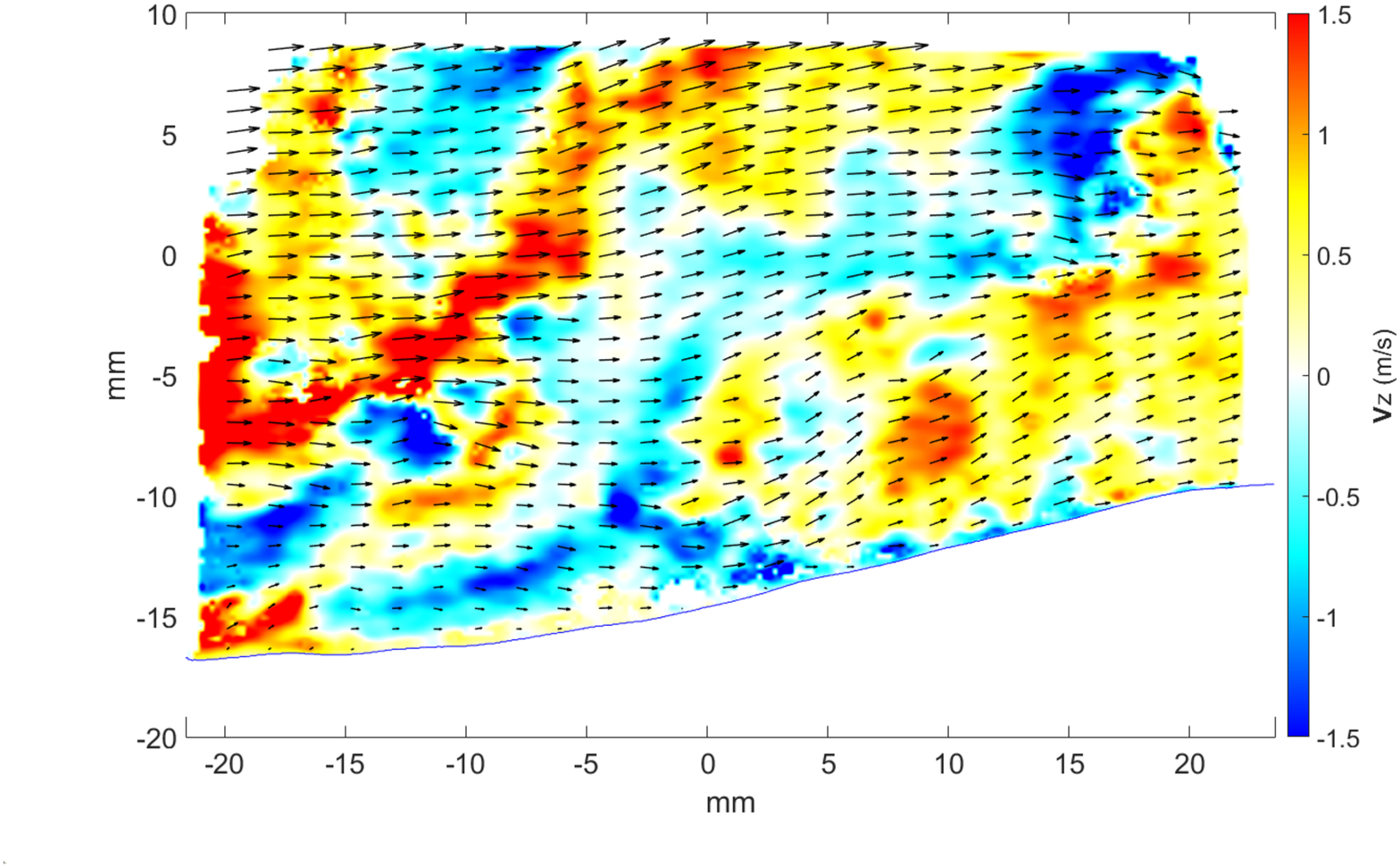}}
\subfigure[]{
\includegraphics[width=0.5\textwidth]{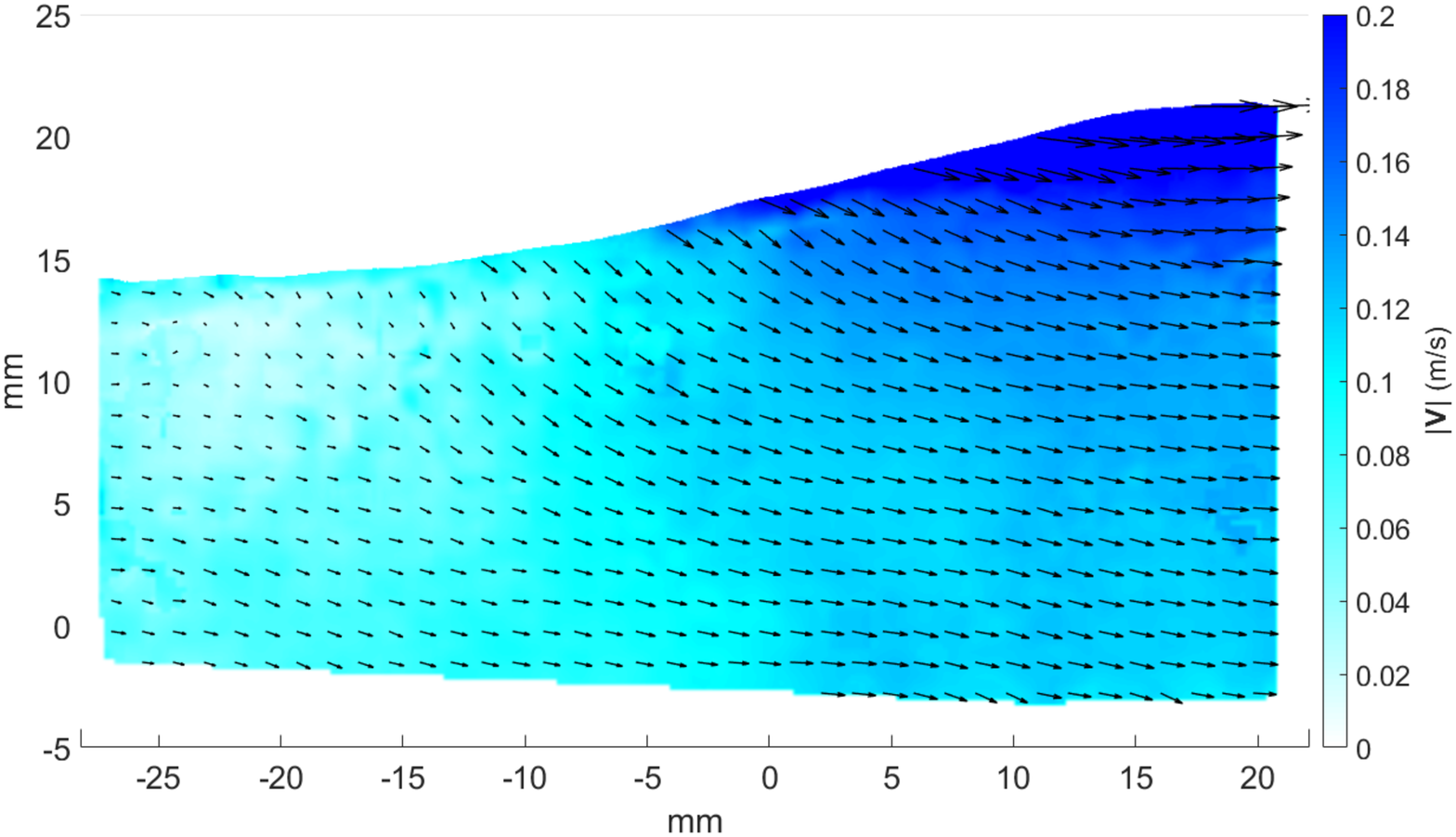}}
\subfigure[]{
\includegraphics[width=0.5\textwidth]{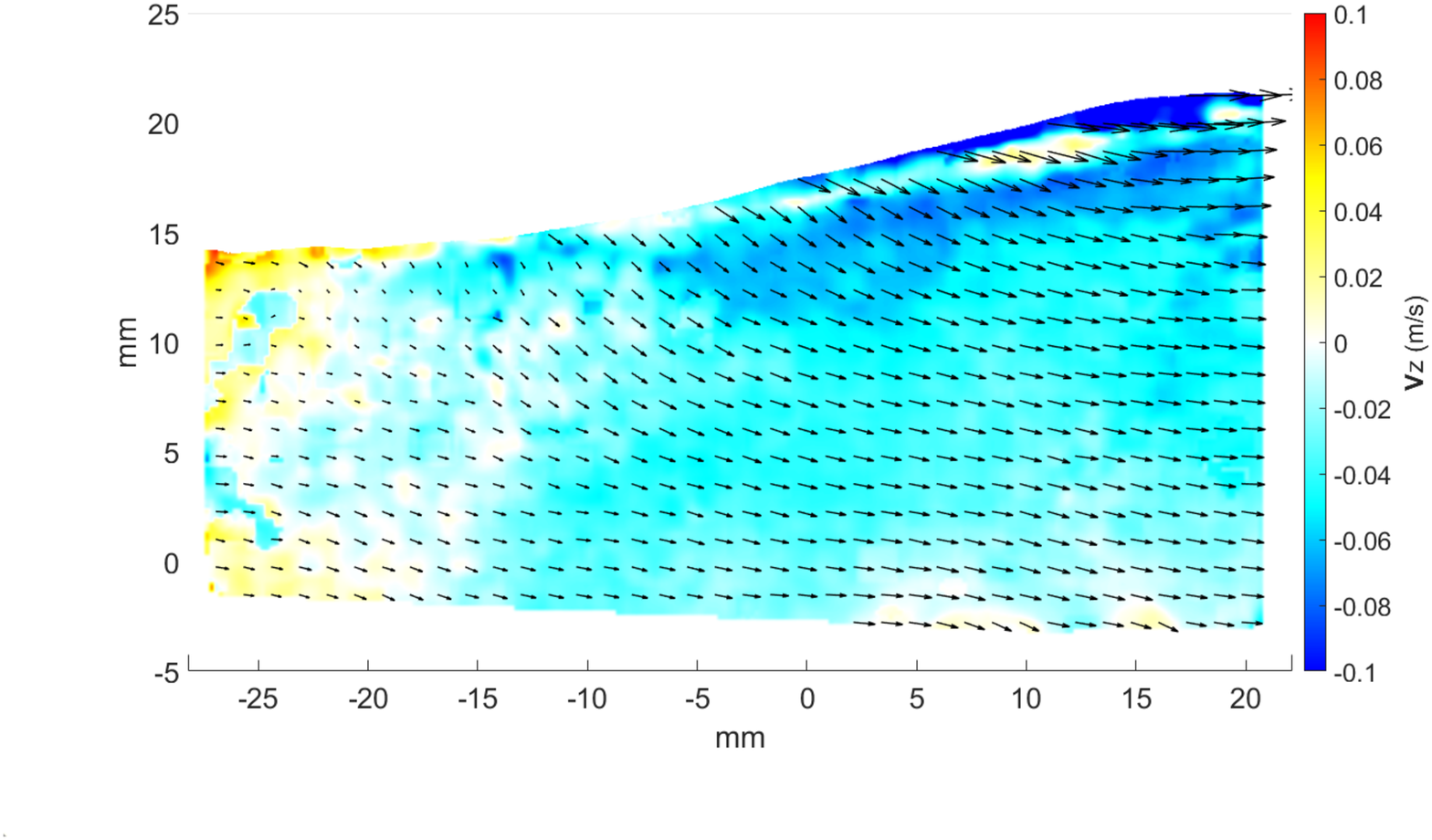}}
\caption{PIV results in region above and below the interface, background in \textbf{a} and \textbf{c} is velocity amplitude and in \textbf{b} and \textbf{d} is z-component of velocity.}
\label{fig:12}
\end{figure*}
\section{Results and discussion}
\label{sec:12}
\begin{figure}
\subfigure[]{
\includegraphics[width=.5\textwidth]{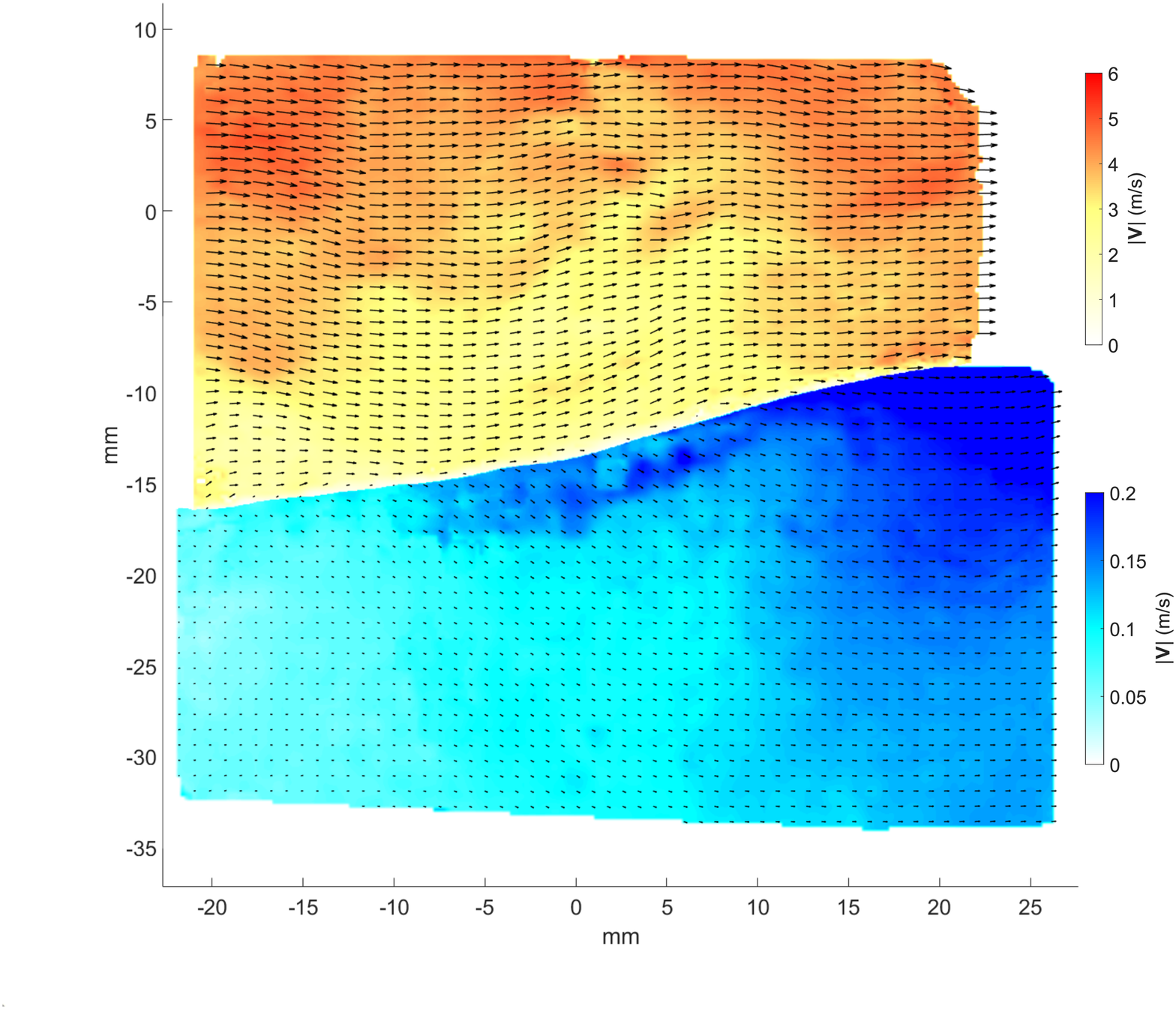}}
\subfigure[]{
\includegraphics[width=.5\textwidth]{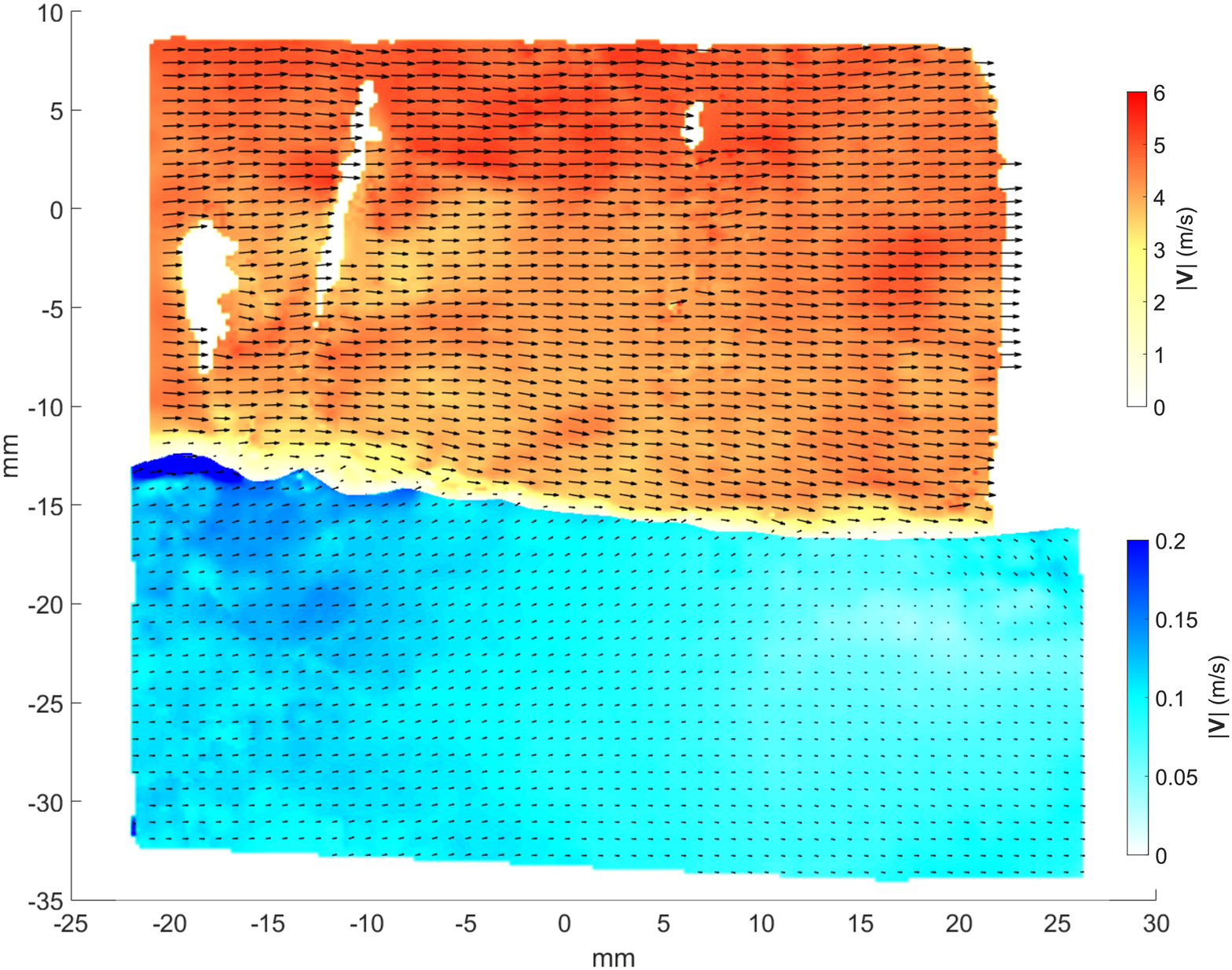}}
\caption{PIV results above the interface and below the interface are shown in one picture, the vector amplification ratio is different in two sides, Vector amplitude is showed in background with warm-toned colour bar and cold-toned colour bar for air flow and water flow.}
\label{fig:13}
\end{figure}

PIV results are shown in Fig. \ref{fig:12} and Fig. \ref{fig:13}, the flow is from left to right in all pictures. Fig. \ref{fig:12} presents vector field above and below the interface in a moment, and its background is the amplitude or z-component of velocity. Fig. \ref{fig:13}a is clip art of vector field above and below the interface, where interaction of wind and current can be seen simultaneously in one picture or video, while Fig. \ref{fig:13}b demonstrates that measurement method in this article also works passable in the region behind wave crest. As the experiment is held in a short wind-water tunnel, the wind-waves have not fully developed. Wind speed is much faster than interface phase speed and interface phase speed is much faster than water flow speed, as shown in Fig. \ref{fig:8} and Fig. \ref{fig:13}. Boundary layer in air flow near interface is very thin, which can be also seen in Fig. \ref{fig:12}, its thickness is about 2 mm at most phase of waves. Despite that, recirculation zone appears on the downstream of crests, which can be seen in Figures and Video. Water flow is similar but not exact Stokes waves especially in the downstream of crests.

\begin{figure}
\includegraphics[width=.5\textwidth]{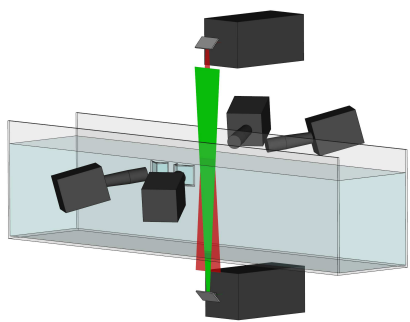}
\caption{A schematic of experimental setup with two laser of different colour.}
\label{fig:14}
\end{figure}
Illumination in air deteriorates when light sheet pass through tiny wavelets on the downstream of wave crests. Due to many convex and concave menisci exist there, light converges or diverges after passing through the interface, which has two unfavourable factors, non-uniform strength of light sheet on FOV and out-of-plane motion of light sheet. As the noise of CMOS is mainly depended on temperature (Lundberg 2002), noise level is roughly similar in the whole frame, signal-to-noise ratio increases in some underexposed region. This has been mended by sliding sum-of-correlation. Besides, heterogeneous illumination $I_{illu}$ in Eq. (\ref{eq:2}) introduces weighting to cross-correlation. As $I_{illu}$ goes with interface moving and $I_{illu}$ goes with particles moving, and the latter is much faster than the former, weighting is nearly unchanged in two adjacent frames, which tends to emphasis on zero displacement. As is known to all, classical stereoscopic PIV uses two cameras and calculates velocity field in the perspective of each camera, then reconstructs three-component velocity field. The final result will not be right as long as one part in reconstruction is error. Fig. \ref{fig:11}c shows the worst situation, PIV without image preprocessing results in many spurious vectors. $I_{illu}$ can be equalized by min-max normalization, actually, normalization with filter length 3, 5 and 7 pixels have been tried, filter of short length adjusts the illumination light to be uniform effectively, but the upper and lower envelops of light intensity in processing change too fast, thus introduce more noise to image, while filter of long length does not introduce so much noise to image, but does not work well when $I_{illu}$ varies too violently. Adaptive min-max normalization is a combination of that with short and long filter length, and produces better results(Fig. \ref{fig:11}d).

Yet, pre- and post-processing can not remedy out-of-plane motion of light sheet. It brings equivalent z-component velocity of particles, we can only remove corresponding spurious vectors. There is a way to completely solve this problem, which is to use two sets of illumination system. See Fig. \ref{fig:14}, red and green (for example) laser illuminate region on one side of interface and filters on cameras can block unwanted light. In present experimental setup with only monochromatic laser, shading light upwards can reduce impact because wind is much faster than interface, illumination changes not so much comparing to particles moving in air.
\section{Conclusions}
A flow visualization method for wind-waves is proposed in this article. Reflection on water side replaces that on air side of interface in flow visualization. In this way, reflected light from interface is recorded with diversion due to the undulation of interface, while refracted light pass through interface will be very weak because of higher refractive index in water and proper exit angle of light. The contrast of photos increases, and more details like tiny wavelets on the downstream of crest are reserved in this minor technical change, which is better than present visualization method and quantitative measurement results. By concatenating visualization photos in different stages from upstream to downstream, the scenery from capillary waves to gravity-capillary waves in the early stage of wind-waves is able to be seen.

As the second topic of this article, a PIV arrangement for wind-waves is introduced, in which flow field above and below the interface can be measured simultaneously by stereoscopic PIV. This arrangement utilized two kinds of tracking particles (smoke and hollow acrylic particles) with different scattering performance to make their size and intensity diverse in photos. The content of photos grabbed by four cameras is analyzed, in the photo of air flow, there are smoke particles above the interface and hollow acrylic particles below the interface, while in the photo of water flow, there are acrylic particles in both side of interface, but particles above the interface are mirror image of interface with higher brightness. Interface detection is first applied on the photos focusing on air flow, then adapted to photos focusing on water flow after spatial and temporal smoothing. Adaptive min-max normalization and near-wall image preprocessing is applied before PIV correlation. The final result of flow field in two phase is shown in one figure in the end.

The imperfection of PIV illumination in experiment is discussed. The quality of light sheet deteriorate after passing through the interface. To radically overcome this problem, it is recommended to make use of two light sources in different colour and illuminate regions on each side of interface individually.

\textbf{Acknowledgments} This research is supported by the National Natural Science Foundation of China under grant numbers 109103010062, 10921202, 11632002, and 11602005.

\end{document}